# Tightening the knot in phytochrome by single molecule atomic force microscopy

Thomas Bornschlögl[1], David M. Anstrom[2], Elisabeth Mey[1], Joachim Dzubiella[1], Matthias Rief[1,3*] and Katrina T. Forest[2,*]

[1]Physics Department, Technische Universität München, D-85748 Garching, Germany,
[2]Bacteriology Department, University of Wisconsin-Madison, Madison, WI 53706
[3]Munich Center for Integrated Protein Science CiPSM, 81377 Munich, Germany

Running title: Single molecule AFM of phytochrome knot

*Address correspondence to: MR, Physik Department E22, Technische Universität München, James-Franck-Straße, D-85748 Garching, Germany, tel. +49 89 28912471, fax +49 89 289 12523, e-mail mrief@ph.tum.de or KTF, 1550 Linden Dr., Madison, WI 53706, tel. 608 265-3566, fax 608 262-9865, e-mail forest@bact.wisc.edu,

Keywords: protein folding; protein structure; *Deinococcus radiodurans*

ABSTRACT A growing number of proteins have been shown to adopt knotted folds. Yet the biological roles and biophysical properties of these knots remain poorly understood. We have used protein engineering and atomic force microscopy to explore single-molecule mechanics of the figure-of-eight knot in the chromophore-binding domain of the red/far red photoreceptor, phytochrome. Under load, apo phytochrome unfolds at forces of ~47 pN, while phytochrome carrying its covalently bound tetrapyrrole chromophore unfolds at ~73 pN. These forces are among the lowest measured in mechanical protein unfolding, hence the presence of the knot does not automatically indicate a super-stable protein. Our experiments reveal a stable intermediate along the mechanical unfolding pathway, reflecting sequential unfolding of two distinct subdomains in phytochrome, potentially the GAF and PAS domains. For the first time, our experiments allow direct determination of knot size under load. In the unfolded chain, the tightened knot is reduced to 17 amino acids, resulting in apparent shortening of the polypeptide chain by 6.2 nm. Steered molecular dynamics simulations corroborate this number. Finally, we found that covalent phytochrome dimers created for these experiments retain characteristic photoreversibility, unexpectedly arguing against dramatic rearrangement of the native GAF dimer interface upon photoconversion.

**Introduction**

For decades it was assumed that proteins do not adopt knotted folds, which is to say that pulling on the C-terminus and N-terminus of a protein would yield an uncomplicated linear polypeptide. However, recent structural biology results (1-3), including the rapid accumulation of three-dimensional coordinates *via* structural genomics efforts and new computational tools for mining the protein data bank for previously unrecognized knots (4-6), have upended this assumption and made clear that some proteins are indeed knotted in the native state. There are now approximately a dozen independent protein folds with recognized knots, not including shallow knots in which only a few residues pass under a loop (7; 8).

Phytochrome is an evolutionarily conserved photoreversible red/far red light photoreceptor with a covalently attached linear tetrapyrrole chromophore. Plant phytochromes mediate germination, flowering, shade avoidance, and senescence among other light responses by trafficking from the cytoplasm to the nucleus upon red (~660 nm) light irradiation (9). Cyanobacterial, eubacterial and fungal phytochromes have been more recently described (10; 11) and thus phytochromes' roles in prokaryotes are less well understood than in plants. *Deinococcus radiodurans* provided the first phytochrome (*Dr*BphP) for which an x-ray crystallographic structure was determined (1). This dark-state Pr structure of the chromophore-binding domain (CBD) of *Dr*BphP comprises an N-terminal extension harboring the covalent attachment site (Cys24) for the biliverdin IXα (BV) chromophore, a PAS domain, and a GAF domain that surrounds the largely buried tetrapyrrole. The native protein contains two additional domains not represented in the CBD, a PHY domain that stabilizes the far-red light absorbing Pfr form of phytochromes, and a histidine kinase domain proposed to mediate the phosphorylation events triggered by red light absorption (9). The *Dr*BphP CBD structure revealed an unanticipated figure-of-eight knotted topology for which there was no prior evidence (Fig. 1a). An ~29 residue insertion between secondary structure elements of a canonical GAF domain forms a lasso for the N-terminal 34 amino acids to create the polypeptide knot (1).

Here, we have investigated the mechanical characteristics of the knot in phytochrome. Forced unfolding of proteins by atomic force microscopy (AFM) has been used to explore the unfolding landscape of proteins, yielding insights into the forces required to unfold a protein, the mapping of folding sub-domains within complicated protein structures, and the free energy landscapes of folding (12-14). AFM is well-suited to the study of mechanical unfolding of knotted proteins, as it provides a way to control the open ends of a protein knot (15; 16). Experimentally, researchers have used single molecule techniques, like optical traps, to construct and manipulate knots for polymeric structures such as DNA helices or actin filaments (27, 28). Ikai and co-workers have reported mechanical unfolding experiments with carbonic anhydrase, containing a shallow knot in its structure. However, those authors found the mechanical stability of this knotted protein was too high to unfold the native state in AFM experiments (17; 18). In this paper we describe the first successful example of single molecule unfolding of an intrinsically knotted protein. Applying single molecule force spectroscopy to the phytochrome CBD figure-of-eight knot, we determined the unfolding force of CBD and the size of the tightly pulled knot in the peptide chain and compared the results with molecular dynamics simulations.

**Results**

*Design and characterization of polymeric phytochromes.* Successful protein unfolding by AFM has relied on natural polymers or on engineered protein interfaces to create covalent homopolymeric proteins. This allows the AFM unfolding event for each domain to be observed multiple times within a single unfolding experiment. We designed phytochrome polymers based on non-native disulfide bonds introduced in positions suggested by close proximity in the crystal packing lattices of *Dr*BphP (1; 19-21). The first disulfide was introduced across the dimer interface at residue 314 with the second at the solvent-accessible residue 18 (Fig. 1A). The double-cysteine substituted variant is a soluble protein which forms covalent dimers and higher-order multimers when oxidized (Fig. 1B). These non-native polymers are expected to consist of alternating head-to-head and tail-to-tail disulfide bridges (Fig. 1C).

The existence of a strong dimer interface mediated by GAF helices in the CBD is itself a relatively new understanding of phytochrome dimerization (19), which has long been suspected to occur predominantly via the histidine kinase domain (19; 22-24). We thus also created the L314C variant in the full-length *Dr*BphP, which became a tool to study the large-scale conformational changes that might take place during photoconversion. This protein readily forms covalent dimers in the absence of reducing agent (Fig. 2). Unexpectedly, these covalent dimers have not only the wild-type Pr spectrum, but they moreover undergo a spectrally normal photocycle in both directions (Fig. 2). This result implies that the orientation of phytochrome remains relatively constant with respect to the CBD dimer interface during photoconversion, both for the red-light driven Pr->Pfr forward reaction and for the dark reversion from Pfr-> Pr.

*Determination of knot size and Dr*BphP CBD-stability*. Force extension curves of the *Dr*BphP CBD polymer with bound BV were measured, and in some cases multiple unfolding peaks were visible indicating the polymer contained several CBD units (Fig. 3A). Worm-like chain (WLC) fits were used to measure the length gain of the polypeptide chain upon unfolding to nm precision (23, 24, 35). In order to get reproducible results, we held the persistence length p = 0.5 nm constant and only fit the contour-length ($L_{18,314}^{Holo}$). The distribution of 47 measured contour-length increases $\Delta L_{18,314}^{Holo}$, has an average value of 98.1 ± 1.0 nm (Fig. 3B). This value is significantly shorter than the expected contour length gain for the transition from a fully folded to a fully extended state. The expected contour length gain can be calculated from formula (1) and is predicted to be $\Delta L_{18,314} = 104.1 \pm 0.6$ nm for the total unfolding of CBD, a value which is larger by 6.2 nm than our experimentally measured one. Hence the stretched polypeptide of *Dr*BphP CBD appears shortened by 17 ± 3 amino acids. A straightforward explanation for this observation is that the tightly pulled knot contains 17 amino acids that make the stretched chain appear shorter than the unknotted chain.

Because the folding pathway *in vivo* for the knotted phytochrome is completely unexplored, it is a formal possibility that the knotting of the protein requires the prior attachment of the tetrapyrrole chromophore. To explore this, we performed the same measurements on apo-phytochrome polymers that had been expressed and purified without addition of biliverdin (Fig. 3C, 3D). The measured contour length increase is 97.5 ± 1.8 nm, identical within error to the holoprotein case.

Although the contour length assigned to the knot is the same for apo *vs.* holo protein, there is a striking and significant difference in the unfolding forces measured in the two experiments; ~73 pN vs. ~47 pN, respectively, for apo and holo CBD (*cf.* Figs. 3A *vs* 3C or Figs. 4C *vs* 4C). The higher forces needed in the presence of chromophore clearly show that BV, covalently attached at residue 24, is buried within the protein structure until the protein reaches the major unfolding peak. Based on this observation, we can exclude the possibility that amino acids following Cys18 have unfolded prematurely at low forces prior to the experimental measurements, as this would have lead to BV being pulled out of the binding pocket. This scenario would have increased the distance between residues 18 and 314 within the monomer, and thus also lead to an apparent shortening of the unfolded polypeptide which we contribute to the tightened knot.

*Unfolding pathway of phytochrome.* Between the major unfolding peaks of both the holo and apo forms of *Dr*BphP CBD, the force-extension traces persistently exhibit a further unfolding event at ~30 pN (Figs. 3A and 3C). These events indicate an intermediate on the mechanical unfolding pathway. The fact that this intermediate is more apparent in apo than in holo traces can be readily explained by the higher unfolding forces of the major peaks for holo-phytochrome, which mask the subsequent low force peaks. We used WLC fits to measure the length of the polypeptide chain unraveled in going from the native to the intermediate state, $\Delta L_I^{Holo}$ (Figs. 4A and 4E). We obtained an average contour-length increase of $\Delta L_I^{Holo} = 68.0 \pm 2.2$ nm for CBD with bound BV (Fig. 4B) and $\Delta L_I^{Apo} = 65.3 \pm 2.6$ nm for the apoprotein (Fig. 4F). These results are identical within error and lead to an estimate that ~180 amino acids unfold in going from the native to the intermediate state. The 186 amino acid GAF-domain is hence the most likely candidate for the domain that unravels first, leaving the remaining 90 amino acid PAS domain to form the unfolding intermediate. This interpretation is corroborated by the sensitivity of the unfolding forces of the major unfolding peak to the presence of chromophore (Figs. 4C and 4G), while the unfolding force distribution for the intermediate peak that marks subsequent unfolding of the PAS domain is independent of chromophore (Figs. 4D and 4H). Because the tetrapyrrole contacts exclusively the GAF domain in the folded protein, the major unfolding peak must be associated with this domain. Moreover, the PAS domain in the bacterial phytochrome shows only few electrostatic and hydrophobic interactions with the GAF domain (1). It is therefore very likely that the sub-domains unfold sequentially, with the GAF domain the first to unfold. The observation that unfolding intermediates follow domain boundaries is reminiscent of the concept of mechanical unfoldons in maltose binding protein, in which mechanical unfolding intermediates formed thermodynamically stable building blocks of the protein (13).

*Validation by molecular dynamics calculations* To corroborate our experimental result of $17 \pm 3$ amino acid residues comprising a taut figure eight knot, we performed steered MD simulations of short $4_1$-knotted polypeptide chains with atomistic resolution of solutes and water. As the final experimental knot location is not known we randomly selected two (30 amino acid long) stretches of phytochrome, which are (I) EPIHIPGSIQ PHGALLTADG HSGEVLQMSL (aa 41-70) and (II) KFAPDATGEV IAEARREGLH AFLGHRFPAS (aa 191-220). Force-extension curves were calculated during peptide pulling, and the contour length differences between the knotted peptide and the unknotted analog were evaluated at relevant pulling forces. At a force of ~70 pN, we find $7.11 \pm 0.4$ nm and $7.25 \pm 0.4$ nm equating to $19 \pm 1$ and $20 \pm 1$ amino acids involved in the tight figure-of-eight knot of peptide I and II, respectively (Fig. 5). This result is within the error of our experimental measurements. The lack of sequence specificity suggests the tightly-pulled knot structure is dominated by packing effects of the backbone, an idea that will be further investigated elsewhere. Interestingly, we observe only a weak force dependence of knot tightening for pulling forces larger than ~70 pN. The increase to a strong force of ~1 nN causes knot lengths to shrink only one amino acid residue further for both peptides, indicating an extremely tight peptide packing and steric hindrance at experimentally and physiologically relevant unfolding forces. It is noteworthy that these tightly-pulled knots are very long-lived; they do not dissolve (unknot) in a ~150 ns free MD simulation.

It is important to note that the sequence-independence of the structure of the tightly pulled knot is not equivalent to a lack of sequence requirement for forming the knot initially. Recent computational work suggests threading in knotted proteins may be an early event in folding and may potentially depend on interactions which are not part of the final folded structure (25).

## Discussion

Very little is yet known about the mechanical properties of protein knots, the mechanisms by which they fold, or the evolutionary advantages they may confer. It has been suggested that protein stability and function are influenced in a beneficial way by the presence of knots (26; 27). Furthermore, it is estimated that by a random distribution, more proteins should be knotted than are observed, rather than less, and based on this it has remained largely unclear why nature steers some proteins to form a complicated knotted structure, while most are discouraged (28; 29). Moreover, a puzzle remains how the energy landscape of a protein fold must be tuned to allow the formation of such a knot in finite time (17, 18). Mallam and Jackson have rigorously shown that the thermodynamic and kinetic unfolding and refolding pathways of the knotted, dimeric methyl transferase YibK are governed by the same parameters known for unknotted proteins (30; 31). The rate limiting steps in its folding are proline isomerization and dimerization, comparable to well-studied unknotted proteins. The fate of the knot itself during folding and unfolding is difficult to address in these studies. One likely possibility is that the formation of the knot is itself an early and non-rate limiting step in protein folding (29).

The functional role of the *Dr*BphP CBD knot has not been determined nor has its folding pathway been examined. However, there exists a widespread class of phytochrome-like proteins that is missing the 29 residue lasso loop (11), and thus cannot be expected to be knotted. These are stable, chromophore-bound photoreversible photoreceptors that can be highly expressed in *E. coli* and purified for biochemical studies (32; 33). Thus, stability of the folded protein alone has not driven the evolution of a knotted phytochrome. Furthermore, a PAS-less and therefore unknotted phytochrome from a thermophilic cyanobacterial species has been characterized (34), and its red far-red photoreversibility is comparable to the longer more well-characterized plant and bacterial proteins. Thus, photoreversibility is also not the sole role of the knot.

The overall mechanical stability of apo CBD is relatively low (~47 pN) compared to other proteins whose unfolding has been studied by single molecule methods (12). Thus, for phytochrome at least, it is difficult to argue that stability is the evolutionary pressure for knot formation. It has also been suggested that some knots might prevent proteins from being translocated inappropriately into the proteasome (7). The radius of gyration data of the tight knot from MD calculations (~0.75 nm) indicates a pore diameter of ~2 nm below which a translocated knotted protein would get stuck. However, eukaryotic phytochrome is degraded by the proteosome upon light activation (35), and bacterial phytochrome would be subject to a different degradation pathway, so this argument also is unlikely to provide the rationale for the evolution of the phytochrome knot (7). We favor another possibility for the knot in phytochrome, which is that it limits the motions the protein domains can undergo relative to one another when light energy is absorbed by the chromophore and used to drive conformational change in the protein (1). Free vibrations would be lost as heat, and unrestricted motions would be inefficient for signal transduction via the multiple domains of phytochrome and its interacting cellular partners.

The constraints imposed by the knot between the PAS and GAF domains may limit motion along the proper trajectory for exposure of appropriate side chains of the GAF domain, for example, to interact productively with the PHY domain and signaling partners downstream of phytochrome itself. Construction of permuted phytochromes lacking the knot should be possible and would be one way to test this hypothesis.

This report paves the way for single molecule unfolding studies of other classes of knots in other knotted proteins, and moreover sets the stage for exciting re-folding experiments (36) in which it should be possible to answer the unresolved question whether the knot forms as a first step in protein folding or a later step. As the number of known protein knots continues to grow, this work becomes generally applicable to a large class of proteins and can reveal fundamental properties about the statistical mechanical behavior of polypeptides.

**Materials and Methods**

*Creating Variant Phytochromes* Disulfide bridge locations were engineered based on visual inspection of crystal packing (20) within two published forms of *Dr*BphP (pdb codes 1ztu and 2o9c) and verification using Disulfide by Design (37).

Beginning with a pET-based overexpression vector (Novagen) encoding the 321 amino acid *Dr*BphP CBD with an N-terminal T7-tag and C-terminal hexahistidine tag (1), rolling circle mutagenesis was carried out in several steps to achieve the plasmid encoding CBD-E18C/L314C. First, CBD-L314C was created using forward primer 5'- CCG CTT GCT GAG CTG TCA AGT TCA GGT CAA GG-3' and reverse primer 5'- CCT TGA CCT GAA CTT GAC AGC TCA GCA AGC GG-3'. The same strategy was also used to create this change in the pET vector that encodes the full-length *Dr*BphP (1). Because the site directed mutation (E18C) was unsuccessful several times, we created an intermediate vector with several silent mutations to remove a predicted eight base pair hairpin surrounding the E18 site (forward primer 5'-GGT GGC CCG GAA ATT ACG ACG GAG AAC TGC GAG CGC-3' and reverse primer 5'-GCG CTC GCA GTT CTC CGT CGT AAT TTC CGG GCC ACC-3'), and then created the desired final gene within this framework (forward primer 5'- GCG TTT ACC TTG GTG GCC CGT GTA TTA CGA CGG AGA ACT GC-3' and reverse primer 5'-GCA GTT CTC CGT CGT AAT ACA CGG GCC ACC AAG GTAAAG CG-3'). All sequences were confirmed by automated DNA sequencing (University of Wisconsin Biotechnology Center).

Expression and purification of full-length and CBD, wild type sequence and variant phytochromes was carried out under green safe lights via Nickel affinity and hydrophobic interaction chromatography as previously described (19), with 1 mM Tris(2-Carboxyethyl) phosphine added during the *in vitro* chromophore ligation step. We found this small change led to a significantly higher fraction of phytochrome containing covalently attached chromophore. For force measurements, protein buffer was exchanged by dialysis to 30 mM Tris pH 8.0, 50 mM NaCl without added reductant and allowed oxidize for several weeks at 4°. In some cases protein crystallization droplets were set up and, although crystals were never observed, this more completely polymerized protein (Fig. 1B) was used for AFM measurements.

*UV/Vis Spectroscopy* Spectroscopy measurements were carried out on a PerkinElmer Life Sciences Lambda 650 spectrophotometer (Waltham, MA). Pr spectra were measured on proteins that had never been illuminated whereas Pfr spectra were measured following saturating irradiation with 690 nm light from a 10-nm half-band width interference filter (Andover Corp. Salem, NH).

*Force Measurements* Single molecule unfolding experiments were performed on a custom-built atomic-force microscope at room temperature. For all measurements we used gold-coated cantilevers (Type B Bio-Lever, Olympus) with spring constants of 6 pN/nm. Before starting the measurement ~20 μl of protein solution (0.5 mg/ml in PBS) were placed upon a clean glass surface and incubated for 5 min. For all force-extension curves, the lever was retracted with constant pulling speeds of 1 μm/s and recorded with 20,000 points/sec.

*Contour Length Calculation* The expected contour length gain upon mechanically unfolding proteins can be calculated according to

$$\Delta L_{i,j} = (j-i)d_{aa} - d_{i,j} \ (1),$$

where $d_{aa}$ is the contour-length of a single amino acid, i and j are the anchoring amino acid positions (18 and 314 in this case), and $d_{i,j}$ is the distance between the anchoring points i and j in the folded structure (38). For a polypeptide persistence length of p= 0.5 nm, $d_{aa}$= 0.365 ± 0.002 nm has been determined to very high precision (23, 24, 35, 40) and $d_{i,j}$ can be measured in the x-ray crystal structure as $d_{18,314}$ = 3.9 nm.

*Molecular Dynamics* We performed standard MD simulations using the software Amber9.0 with the ff03 force-field (39) at fixed pressure (P=1 bar) and temperature (T=300 K) involving ~8000 atoms in a periodically repeated and anisotropic simulation box with Ewald electrostatics. Polypeptides were generated using the Amber tleap tool. Figure-of-eight knots were tied into them by the user, utilizing interactive MD (IMD) in VMD (40). Thereafter, the system was equilibrated for ~5 ns with Langevin dynamics, solvated with TIP3P water and neutralizing Na$^+$-ions, and further equilibrated by a ~5 ns MD simulation. For peptide pulling we utilized the Amber steered MD (SMD) tool with a constant velocity of 0.01 nm/ns to drive the first and last atom of the backbone in opposite directions. Knot lengths and their errors were estimated from the calculated force-extension *F(l)* curves and their fluctutations, respectively. The average force *F* was found to increase monotonically with extension *l*. At the experimentally relevant force of ~70 pN we measured an amino acid contour length of 0.365 ± 0.001 in accord with experiments.

**Acknowledgements** This work was supported by the National Science Foundation (Grant MCB-0424062 to KTF and Richard Vierstra) and the Deutsche Forschungsgemeinschaft (Grant RI 99013-1 to MR and support from the Emmy-Noether-Program to JD). Computing time on the HLRB11 was provided by the LRZ Munich. Dr. J.R. Wagner created the *Dr*Bph CBD L314C-encoding pET plasmid. KTF would like to acknowledge the Max Planck Institut für Plasma Physik for hosting her as a visiting scientist during the time this project was initiated.

**Figure Legends**

FIGURE 1. Design and characterization of phytochrome polymers. A. Cysteines were engineered at positions 18 and 314 (yellow spheres) within the CBD of *Dr*BphP (N-terminus blue with reverse rainbow coloring to red at the C-terminus). Boundaries of the PAS (licorice representation only) and GAF (includes ribbon representation) domains are marked by ovals. Residue L314 is part of the GAF domain dimer interface observed in all crystal forms to date, whereas residue E18 is near a 2-fold symmetry axis in the $P2_12_12$ crystal form (pdb code 2o9c) (1). B. *Dr*BphP CBD-L314C/E18C oligomerizes in a redox-dependent manner. Lanes 1: Holoprotein 2: Holoprotein+DTT 3: Apoprotein 4: Apoprotein + DTT 5: Molecular Weight Markers 6: Pseudo-crystallized CBD polymers. The calculated mass of the expressed monomer is 37 kDa. C. Schematic of *Dr*BphP CBD E18C/L314C polymerized in a head-tail:tail-head:head-tail:tail-head arrangement (engineered disulfide bridges as yellow bars). The simplified subunit knot coordinates were generated using the Knot Server (6) with *Dr*BphP coordinates 3o9c.pdb (19) (reverse rainbow color scheme as in A).

FIGURE 2. Covalently dimerized full-length *Dr*BphP L314C undergoes normal photoconversion upon illumination with 690 nm light (dashed line, *Dr*BphP WT; solid line, *Dr*BphP 314C). Inset: SDS-PAGE gel of illuminated proteins showing Lanes 1: WT+DTT 2: WT 3: L314C+DTT 4: L314C. Pre-illuminated and dark-reverted proteins have the same gel profile (not shown).

FIGURE 3. AFM unfolding of *Dr*BphP CBD (A) Force-extension trace (red) of a single CBD-polyprotein with bound chromophore. Theoretical fits are according to the WLC model (41) (blue). (B) Histogram of contour-length increases $\Delta L_{18,314}^{Holo}$ caused by total unfolding of CBD units with bound chromophore. Vertical black line at 104 nm represents the calculated contour-length for an unknotted CBD. (C) Force-extension trace of a single apo *Dr*CBD-polyprotein. (D) Histogram of contour-length increases for the total unfolding of the apo-form.

FIGURE 4. Unfolding Pathway of *Dr*BphP CBD. (A) Zoom into a single holo-CBD unfolding event (framed in Fig 2A). The rising force connects to a stable intermediate state, with WLC-fit shown as green line. (B) Histogram of contour-length increases from the totally folded to the intermediate state $\Delta L_I^{Holo}$. (C) Histogram of forces needed to unfold the first part of the structure (red circles in A). Average unfolding forces are $F_{18,314}^{Holo} = 73.3 \pm 6.6$ pN. (D) Histogram of forces $F_I^{Holo}$ needed to unfold the intermediate state to the fully extended protein (grey circles in A). The average unfolding forces are $F_I^{Holo} = 32.8 \pm 4$ pN. (E) Zoom into a single apo-CBD unfolding event (framed in Fig 2C). $\Delta L_I^{Apo}$ as fit by WLC is shown as green line. (F) Histogram of $\Delta L_I^{Apo}$. (G) Histogram of forces needed to unfold the first part of the structure (red circles in E). The average unfolding force is $F_{18,314}^{Apo} = 47.1 \pm 3.2$ pN. h) Histogram of forces needed to unfold the intermediate state (grey circles in E), with an average value of $F_I^{Apo} = 27.7 \pm 2.8$ pN.

FIGURE 5. An MD snapshot of the tightly pulled peptide II (*Dr*BphP residues 191-220).

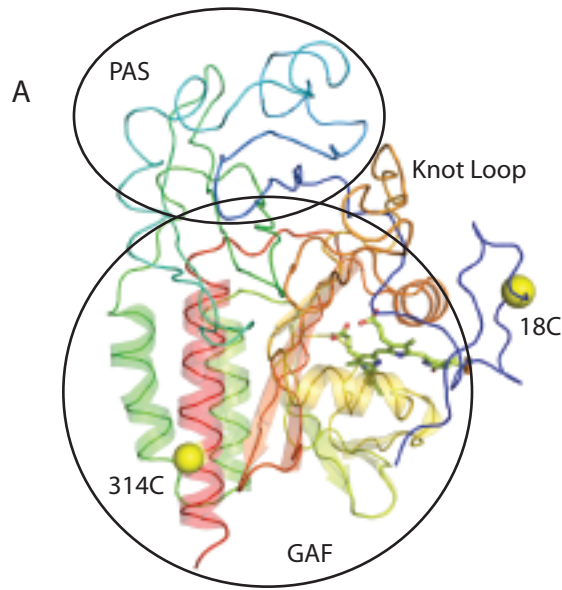

A

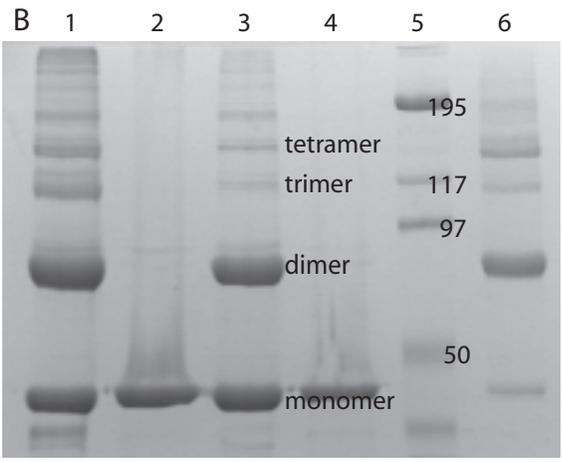

B

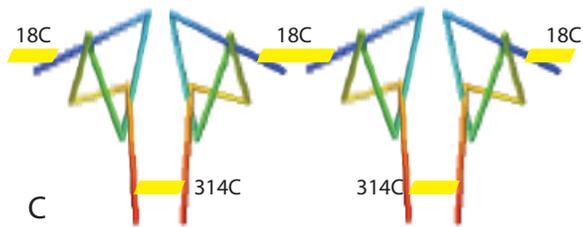

C

Bornschlögl *et al.*, Fig. 1

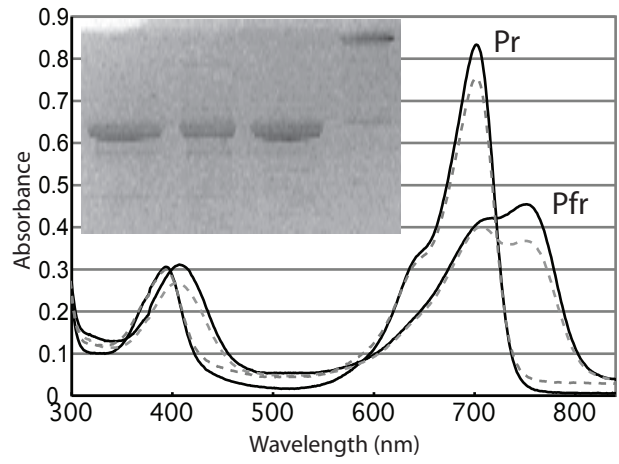

Bornschlögl, *et al*. Fig. 2

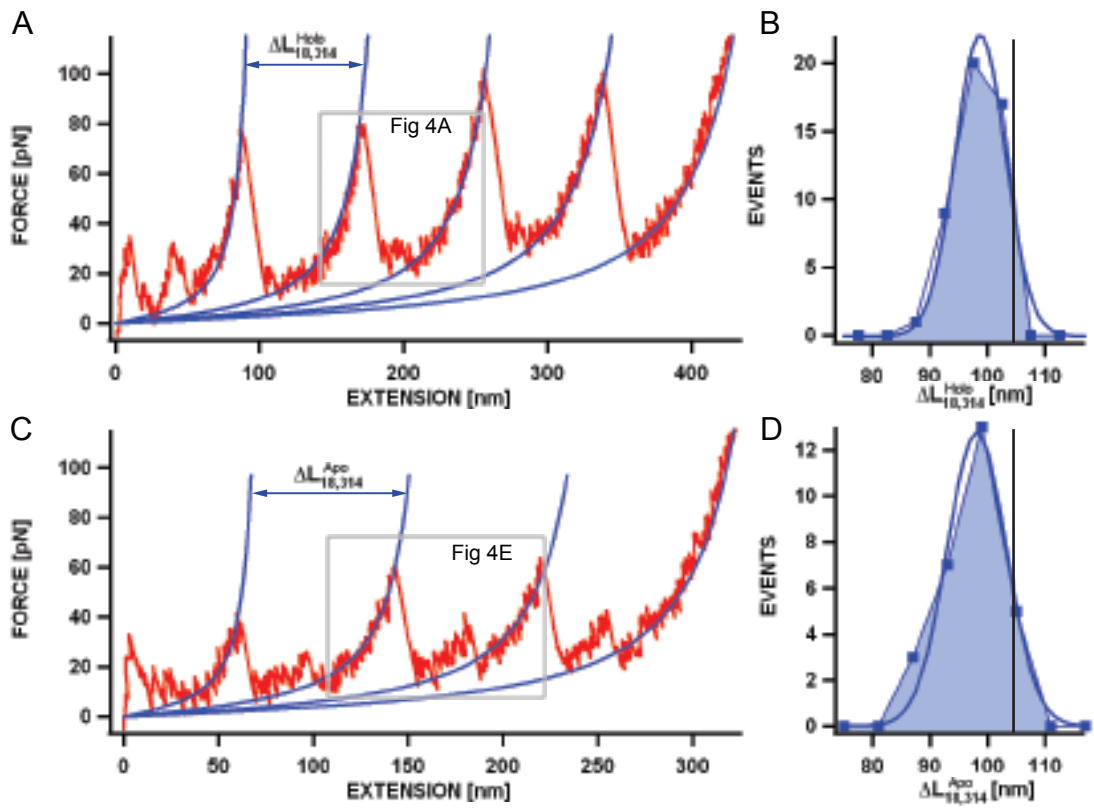

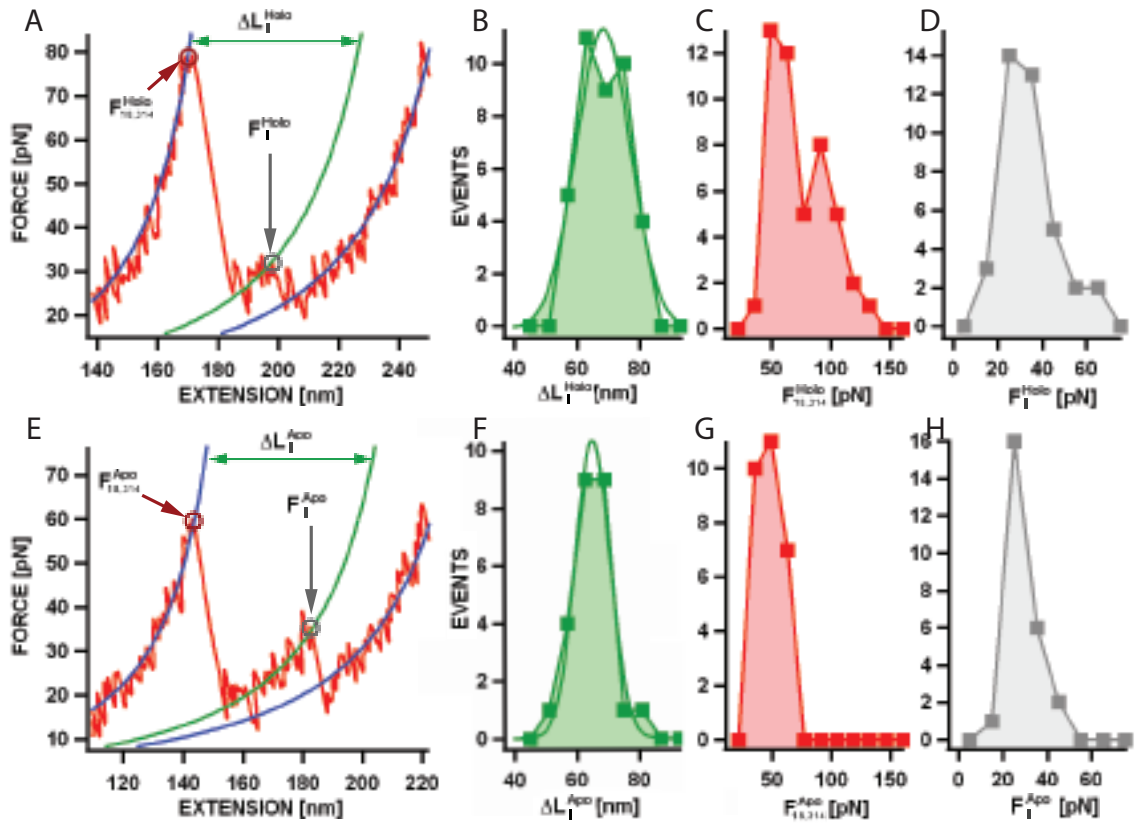

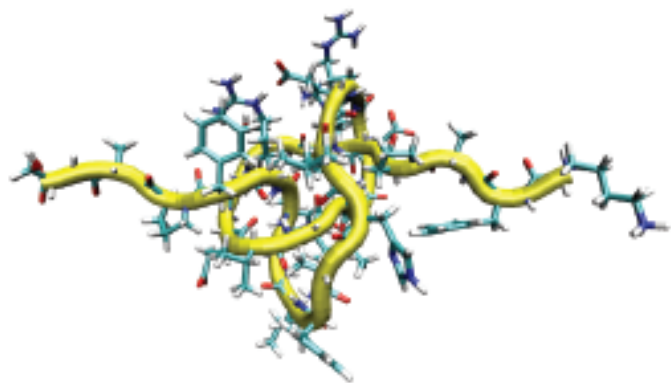

Bornschlögl, *et al*. Fig. 5